\documentclass[a4paper,11pt]{article}
\usepackage{pos}
\usepackage{slashed}

\title{$D^0\overline{D^0}$ mixing from nonlocal condensate contributions}

\author[a]{Lovro Dulibić}
\author*[a]{Blaženka Melić}
\author[b]{Alexey A. Petrov}

\affiliation[a]{Ruđer Bošković Institute,\\
  Bijenička cesta 32, Zagreb, Croatia}

\affiliation[b]{Department of Physics and Astronomy, University of South Carolina,\\
712 Main St., Columbia, South Carolina, USA}

\emailAdd{ldulibic@irb.hr}
\emailAdd{melic@irb.hr}
\emailAdd{apetrov@sc.edu}

\abstract{A significant discrepancy, spanning multiple orders of magnitude, exists between the leading order contribution to the $D^0\overline{D^0}$ mixing parameters and experimental values. This is largely due to the Glashow–Iliopoulos–Maiani (GIM) mechanism, which results in substantial suppression of the theoretical predictions. To bridge this gap, various efforts have been made to account for higher-order terms and nonperturbative effects, which, although suppressed in the operator product expansion (OPE), could potentially lead to a larger contribution by weakening the GIM cancellation through flavour SU(3) symmetry breaking. In this work, we compute the long-distance contributions of nonlocal QCD condensates within different models and, for the first time, determine the impact of the mixed condensate. Although our results still fall short of the experimental value, they represent an improvement over current theoretical estimates by an order of magnitude.}

\FullConference{%
  42nd International Conference on High Energy Physics,\\
  18-24 July 2024\\
  Prague, Czech Republic
}

\definecolor{blue}{rgb}{0,0,0.5}
\definecolor{darkgreen}{RGB}{0,175,10}
\definecolor{brown}{RGB}{150,50,0}

\begin{document}
\maketitle

\section{Introduction}

Most recent HFLAV results (September 2023) \cite{HFLAV:2022pwe} found the average experimental value for the mass difference mixing parameter of neutral charm mesons to be

\begin{equation}\label{eq:experiment}
     x_D= \frac{\Delta M}{\Gamma_D} = (0.407 \pm 0.044)\%\,.
\end{equation}
On the theory side, a calculation taking into account short-distance leading and next-to-leading perturbative terms, gives a result which is orders of magnitude away from the experimental value \cite{Golowich:2005pt}. 
This is due to strong CKM suppression and very efficient GIM cancellation in the box-diagrams leading to the mixing, although separate contributions are large. 
It was demonstrated in \cite{Lenz:2020efu} that the issue could be resolved by selecting the renormalization scale for each internal quark contribution individually. In contrast, the exclusive approach \cite{Falk:2001hx} and the dispersive approach \cite{Li:2020xrz,Li:2022jxc,Falk:2004wg} showed that the experimental value can be achieved by accounting for SU(3) breaking effects. However, these approaches suffer from large uncertainties and uncontrollable theoretical errors.


Already a long time ago there was an idea \cite{Georgi:1992as,Ohl:1992sr} that the inclusion of contributions from higher-order operators in HQET, although suppressed, may lead to a less effective GIM cancellation, giving rise to a larger final value for mixing in the charm system. In addition, because of the not-so-large charm-quark mass, the long-distance, non-perturbative effects cannot be ignored in charm-meson mixing. This, however, questions how fast the OPE expansion would converge.

In this talk, in the inclusive approach, we focus on the long-distance contribution to charm meson mixing which lifts the GIM suppression via chirality breaking terms, specifically arising from \textit{QCD condensates}. The basic idea was first promoted in \cite{Bigi:2000wn}. 
The QCD condensates have been known for some time and appear regularly in the QCD sum rules method (\cite{pascual2014qcd} and others) to compute various non-perturbative quantities including hadronic decay constants and transition form factors. To our knowledge, \textit{nonlocal} condensates (NLC) were initially applied to assess the pion distribution amplitude, aiming to  enhance the precision of `traditional' sum rules with local condensates in \cite{Mikhailov:1986be}. However, the first comprehensive exploration of condensate nonlocality occurred earlier, particularly within heavy-quark propagators and their mass expansions \cite{Gromes:1982su,Generalis:1983hb,Yndurain:1989hp, Bagan:1985zp, Bagan:1993by, Jamin:1992se}. 

An analysis of local vacuum condensate contributions to charm mixing was carried out by \cite{Bobrowski:2012jf}, focusing on the leading chirality breaking term of the diquark condensate. We extend the study of their (unpublished) work by incorporating models for the nonlocal QCD condensate expansion, and also including  higher-dimensional condensates (both mixed and four-quark), which may provide significant contributions. Finally, by introducing the nonlocality of the condensates, we probe the convergence of $1/m_c$ expansion.

\section{Nonlocal condensates}
Typically, the leading order contribution originates from the so-called box diagram. The propagators of internal quarks lie between weak interaction currents $(V-A)$, allowing only structures with an odd number of gamma matrices to contribute. 
The GIM mechanism implies that any terms independent of the internal quark masses will cancel out. Due to the left-handed nature of weak interactions, each propagator in the box diagram contributes two powers of the quark mass. As a result, the box diagram is proportional to ($m_s/m_c)^4$. However, when considering the contributions from background quark condensates, the condensate flips helicity and the suppression is reduced to $\propto \left( m_s/m_c \right)^3$.

The nonlocal quark-quark condensate can be expressed through an expansion in terms of local condensates, as can be found in  \cite{pascual2014qcd,Grozin:1994hd}: 
\begin{equation}\label{eq:qq-exp}
\begin{aligned}
    \langle \overline{q}(x)^a_\alpha q(0)_\beta^b\rangle=\frac{\langle\overline{q}q\rangle}{4N_C}\delta^{ab}&\Bigg[ \delta_{\alpha\beta}\left( 1
    -\frac{x^2}{4}\left(\frac{m^2}{2}-\frac{\langle\overline{q}ig\sigma G q\rangle}{4\langle\overline{q}q\rangle} \right) \ldots \right)+ \\
    +i(\slashed{x})_{\beta\alpha}&\left( \frac{m}{4} - \frac{x^2}{4}\left(\frac{m^3}{12} -\frac{m}{12}\frac{\langle\overline{q}ig\sigma G q\rangle}{2\langle\overline{q}q\rangle}+\frac{2}{81}\pi \alpha_s \frac{\langle\overline{q}q\rangle^2}{\langle\overline{q}q\rangle}\right)\ldots\right)\Bigg],
    \end{aligned}
\end{equation}
where $a,\,b$ are color indices, $\alpha,\,\beta$ are Dirac indices. Local condensates are simplified as $\langle\overline{q}(0)q(0)\rangle=\langle\overline{q}q\rangle$, similarly for the  mixed condensate. The two terms with distinct Dirac structures can be resummed into functions that depend on $x$ and the condensate values. The idea was presented in \cite{mikhailov_nonlocalQCD1992}, where the regular large $x$ behaviour of the expansion was questioned. This led to the use of models incorporating a nonlocality parameter $\lambda_q^2\propto\langle\overline{q}D^2q\rangle \propto \langle\overline{q}ig\sigma G q\rangle$ to regulate the large $x$ behaviour. The coefficients in these models are adjusted to match  the expansion in the local limit $\lambda_q^2\to 0$ by imposing normalisation conditions on the functions that model the condensates. 
By substituting these general functions for the expansion, we can directly work with the non-local condensate, instead of dealing with its local expansion. This enables us to control large $x$ behaviour effectively,  while correctly reproducing the expansion for small $x$: 
\begin{equation}\label{eq:qqNLCgeneralF}
    \langle \overline{q}(x)^a_\alpha q(0)_\beta^b\rangle=\frac{\langle\overline{q}q\rangle}{4N_C}\delta^{ab}  \bigg[ \delta_{\alpha\beta} F_S(x)+i(\slashed{x})_{\beta\alpha}F_V(x) \bigg ]\,, 
    \end{equation}

To achieve the correct order of expansion of the diagrams 
in a background field, we likewise expand the non-local mixed condensate to the same order, 
\begin{equation}\label{eq:GqqExpansion}
\begin{aligned}
   \langle\overline{q}^{a}_{\alpha}(x)G_{\mu\nu}^{cd}(0)q_{\beta}^{b}(0)\rangle=\frac{4\langle\overline{q}q\rangle}{1536}\left(\delta^{bd}\delta^{ac}-\frac{1}{3}\delta^{cd}\delta^{ab}\right)\Bigg[ \big( \sigma_{\mu\nu}+&\\
   +\frac{m}{2}\left(i\sigma_{\mu\nu}\slashed{x}+\gamma_\mu x_\nu-\gamma_\nu x_\mu \right)\big)_{\beta\alpha}\frac{\langle\overline{q}ig\sigma G q\rangle}{\langle\overline{q}q\rangle}+\frac{i}{2}\left(\slashed{x}\sigma_{\mu\nu}\right)_{\beta\alpha}&\frac{16}{9}\pi\alpha_s^{NP}\frac{\langle\overline{q}q\rangle^2}{\langle\overline{q}q\rangle}+\ldots \Bigg]
\end{aligned}
\end{equation}
and modify it in a similar manner using a general function $F_G(x)$, which models large $x$ behaviour\footnote{Here, the Dirac structure is more complex. Ideally, each Dirac structure would have its own higher-order corrections (similar to the quark-quark condensate). Nonetheless, since we are considering, at most, the leading order of the mixed condensate contribution, we are justified in using a single general function $F_G(x)$ to model the behaviour for large $x$.} as 
\begin{equation}\label{eq:GqqNLCgeneralF}
\begin{aligned}
    \langle\overline{q}^{a}_{\alpha}(x)G_{\mu\nu}^{cd}(0)q_{\beta}^{b}(0)\rangle=\frac{8\langle\overline{q}q\rangle}{1536}\left(\delta^{bd}\delta^{ac}-\frac{1}{3}\delta^{cd}\delta^{ab}\right)\Bigg[ \big( \sigma_{\mu\nu}+&\\
   +\frac{m}{2}\left(i\sigma_{\mu\nu}\slashed{x}+\gamma_\mu x_\nu-\gamma_\nu x_\mu \right)\big)_{\beta\alpha}\lambda_q^2+\frac{i}{4}\left(\slashed{x}\sigma_{\mu\nu}\right)_{\beta\alpha}&\frac{16}{9}\pi\alpha_s^{NP}\frac{\langle\overline{q}q\rangle^2}{\langle\overline{q}q\rangle}\Bigg]F_G(x)\,.
\end{aligned}   
\end{equation}
The primary parameters in our nonlocal model analysis include $\lambda_q^2 = 0.4 \pm 0.1$ GeV$^2$, the value of the quark condensate $\langle \bar{q}q \rangle = -(243\,\mathrm{MeV})^3$, and the ratio of the $s$-quark to the light quark condensate $\langle\overline{s}s\rangle/\langle\overline{q}q\rangle=0.8\pm0.3$.

The general functions $F_V(x),\,F_S(x),\,F_G(x)$ introduced above represent a partial resummation of the OPE to all orders.  Their small $x$ behavior is tightly constrained by the requirement to match the expansions, leading to specific coefficients of the general functions that are determined through this matching process. This is essential for ensuring consistency of the approach. However, we have some freedom in modeling their large $x$ behaviour. 
The `delta' model, introduced in references like \cite{Mikhailov:1986be, Mikhailov:1988nz}, utilizes delta functions to fix the coefficients to the expansion, simplifying the representation. 
On the other hand, the more complex ansatz utilized in works of \cite{Braun:1994jq, Braun:2003wx} introduces a large $x$
behavior characterized by an exponential decay, specifically 
$e^{-\lambda_q \sqrt{-x^2}}$. This form aligns with expectations from QCD, where correlation functions typically exhibit rapid decay at large distances, reflecting the confinement of quarks and gluons.
Although both models are constrained to match the small $x$
expansion, the disparity in their large
$x$ behavior could lead to notable differences in predictions. 

\begin{figure}
    \centering
    \includegraphics[width=0.7\linewidth]{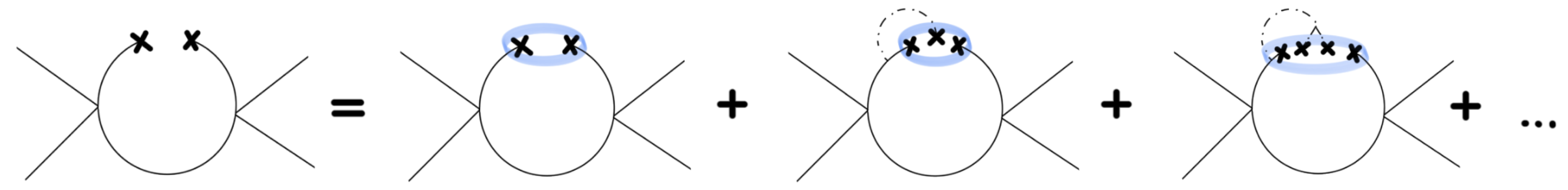}\\
    \hspace{-1.5cm}\includegraphics[width=0.6\linewidth]{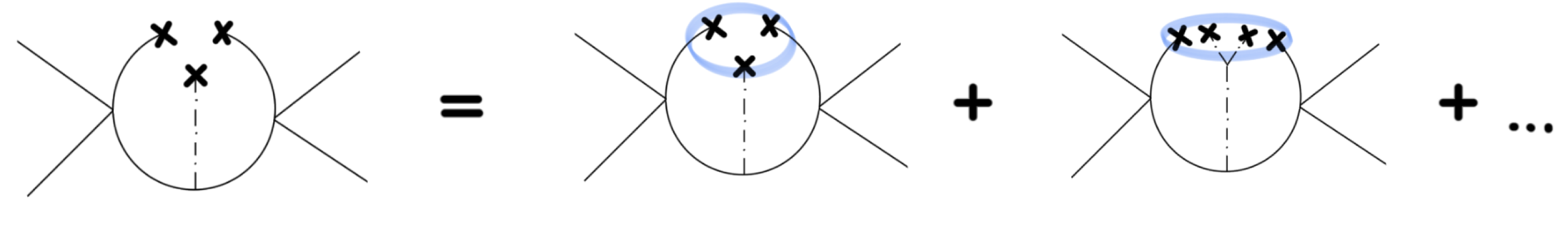}\\
    \hspace{-4.6cm}\includegraphics[width=0.4\linewidth]{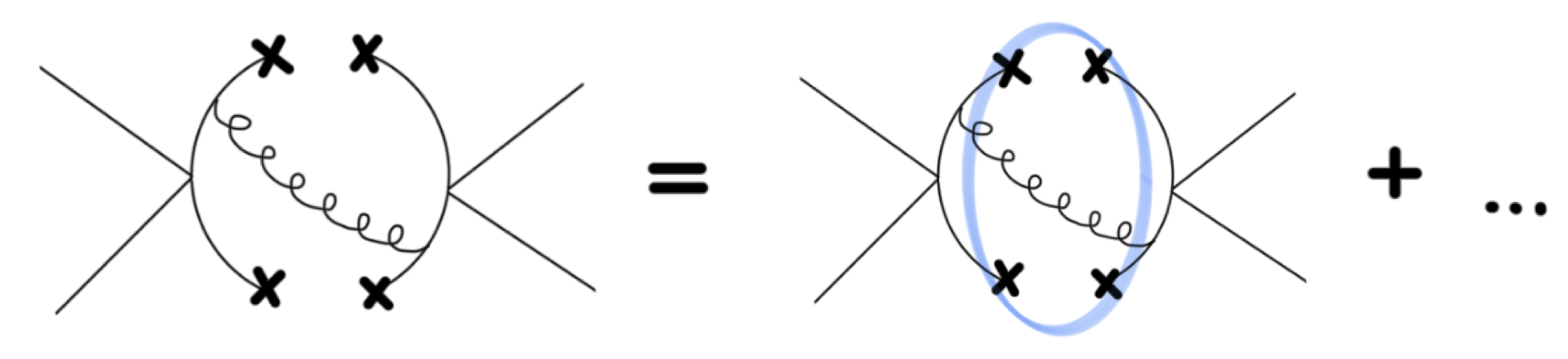}
    \caption{Diagrammatic view of the local expansion of nonlocal condensates: the quark condensate \eqref{eq:qq-exp}, the mixed condensate \eqref{eq:GqqExpansion}, and the four-quark condensate, respectively, up to the same order $\propto \langle \bar{q} q \rangle^2$. The dashed-line represents the background soft gluon field. Note that the genuine four-quark contributions necessarily contain a perturbative gluon exchange.}
    \label{fig:enter-label}
\end{figure}

\section{Results \& outlook} 

Using the simplest delta model, we have calculated the contributions of the quark condensate and the mixed quark gluon condensate, yielding preliminary results of  
\begin{equation}
    x_D^{NLC}=(5.8+1.9)\times 10^{-6}=7.7\times 10^{-6},
\end{equation}
where the two terms represent contributions of the quark and mixed condensates, respectively. In contrast, the LO and NLO short-distance calculations from \cite{Golowich:2005pt} estimate $x_D^{(LO+NLO)}\approx 6\times 10^{-7}$, indicating that our result represents an improvement by an order of magnitude.

\begin{figure}
    \centering
    \includegraphics[width=0.6\linewidth]{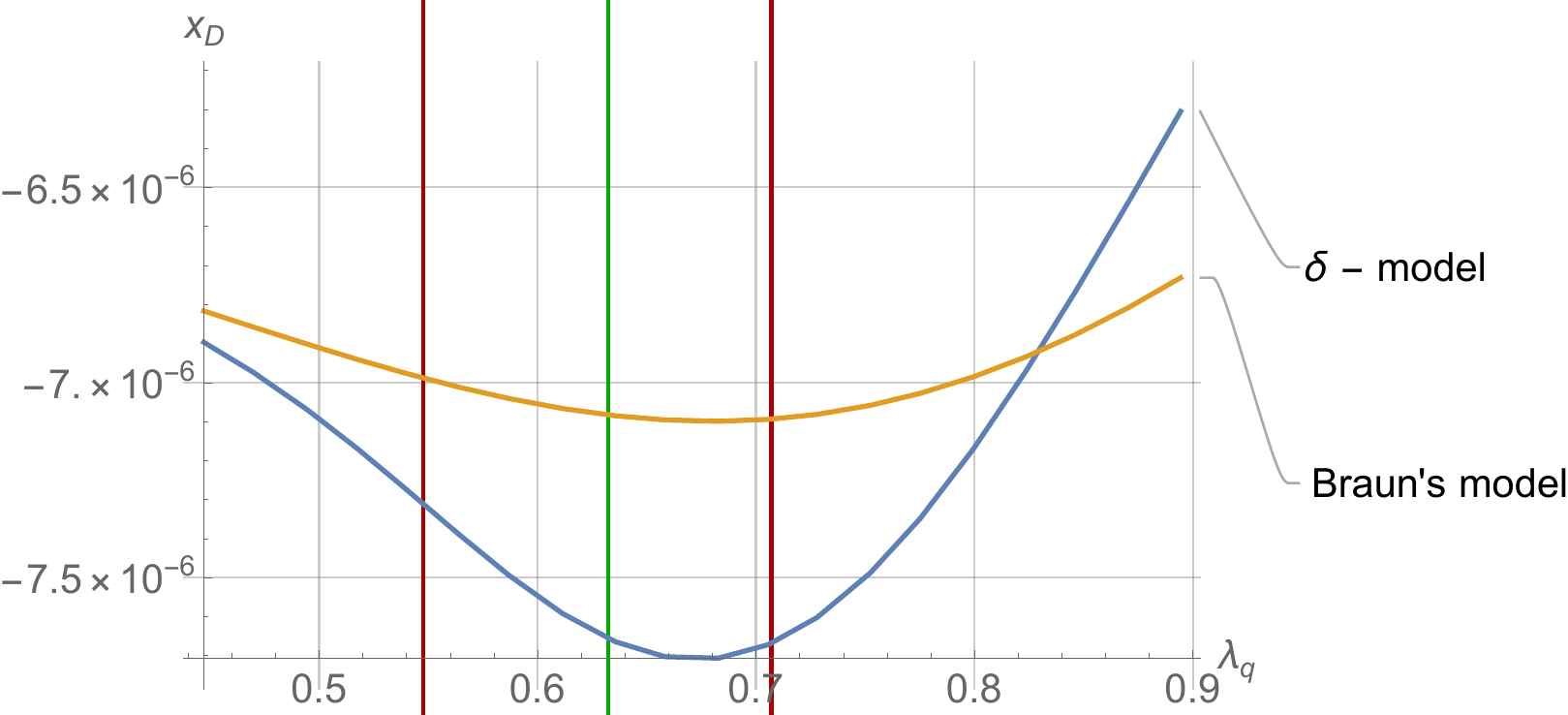}
    \caption{Dependence of the result on the quark virtuality $\lambda_q$ for the `delta' model \cite{Mikhailov:1986be,Mikhailov:1988nz} and `Braun's model' from \cite{Braun:1994jq,Braun:2003wx}. }
    
    \label{fig:lambda dependance}
\end{figure}

Comparing the different models discussed earlier (see Fig. \ref{fig:lambda dependance}), we observe that a more complex model does not lead to significantly different results, but it demonstrates greater stability with respect to the nonlocality parameter $\lambda_q^2$.

In conclusion, we have investigated the contributions of non-local condensates to the mixing of neutral charmed mesons. Our preliminary results indicate that by modeling nonperturbative effects with NLCs, the GIM suppression of the mass difference mixing parameter can be alleviated, although it still falls short of matching the experimental value. For the first time, we have also considered the contribution from mixed condensates and plan to conduct a comprehensive NLC calculation that includes the four-quark condensate contribution. Its expected dependence on the strange mass, $O(m_s^2/m_c^2)$, could significantly impact the mixing of charm mesons. The results will be presented in a forthcoming publication \cite{Dulibicetal}.

\bibliography{refs.bib}

\providecommand{\noopsort}[1]{}\providecommand{\singleletter}[1]{#1}%

\providecommand{\href}[2]{#2}\begingroup\raggedright\begin{thebibliography}{10}

\bibitem{HFLAV:2022pwe}
Y.~Amhis et~al., \emph{{Averages of $b$-hadron, $c$-hadron, and $\tau$-lepton properties as of 2021}}, \href{https://doi.org/10.1103/PhysRevD.107.052008}{\emph{Phys. Rev. D} {\bfseries 107} (2023) 052008}, [\href{https://arxiv.org/abs/2206.07501}{{\ttfamily 2206.07501}}].

\bibitem{Golowich:2005pt}
E.~Golowich and A.~A. Petrov, \emph{{Short distance analysis of D0 - anti-D0 mixing}}, \href{https://doi.org/10.1016/j.physletb.2005.08.023}{\emph{Phys. Lett. B} {\bfseries 625} (2005) 53--62}, [\href{https://arxiv.org/abs/hep-ph/0506185}{{\ttfamily hep-ph/0506185}}].

\bibitem{Lenz:2020efu}
A.~Lenz, M.~L. Piscopo and C.~Vlahos, \emph{{Renormalization scale setting for D-meson mixing}}, \href{https://doi.org/10.1103/PhysRevD.102.093002}{\emph{Phys. Rev. D} {\bfseries 102} (2020) 093002}, [\href{https://arxiv.org/abs/2007.03022}{{\ttfamily 2007.03022}}].

\bibitem{Falk:2001hx}
A.~F. Falk, Y.~Grossman, Z.~Ligeti and A.~A. Petrov, \emph{{SU(3) breaking and D0 - anti-D0 mixing}}, \href{https://doi.org/10.1103/PhysRevD.65.054034}{\emph{Phys. Rev. D} {\bfseries 65} (2002) 054034}, [\href{https://arxiv.org/abs/hep-ph/0110317}{{\ttfamily hep-ph/0110317}}].

\bibitem{Li:2020xrz}
H.-N. Li, H.~Umeeda, F.~Xu and F.-S. Yu, \emph{{$D$ meson mixing as an inverse problem}}, \href{https://doi.org/10.1016/j.physletb.2020.135802}{\emph{Phys. Lett. B} {\bfseries 810} (2020) 135802}, [\href{https://arxiv.org/abs/2001.04079}{{\ttfamily 2001.04079}}].

\bibitem{Li:2022jxc}
H.-n. Li, \emph{{Dispersive analysis of neutral meson mixing}}, \href{https://doi.org/10.1103/PhysRevD.107.054023}{\emph{Phys. Rev. D} {\bfseries 107} (2023) 054023}, [\href{https://arxiv.org/abs/2208.14798}{{\ttfamily 2208.14798}}].

\bibitem{Falk:2004wg}
A.~F. Falk, Y.~Grossman, Z.~Ligeti, Y.~Nir and A.~A. Petrov, \emph{{The D0 - anti-D0 mass difference from a dispersion relation}}, \href{https://doi.org/10.1103/PhysRevD.69.114021}{\emph{Phys. Rev. D} {\bfseries 69} (2004) 114021}, [\href{https://arxiv.org/abs/hep-ph/0402204}{{\ttfamily hep-ph/0402204}}].

\bibitem{Georgi:1992as}
H.~Georgi, \emph{{D - anti-D mixing in heavy quark effective field theory}}, \href{https://doi.org/10.1016/0370-2693(92)91274-D}{\emph{Phys. Lett. B} {\bfseries 297} (1992) 353--357}, [\href{https://arxiv.org/abs/hep-ph/9209291}{{\ttfamily hep-ph/9209291}}].

\bibitem{Ohl:1992sr}
T.~Ohl, G.~Ricciardi and E.~H. Simmons, \emph{{D - anti-D mixing in heavy quark effective field theory: The Sequel}}, \href{https://doi.org/10.1016/0550-3213(93)90364-U}{\emph{Nucl. Phys. B} {\bfseries 403} (1993) 605--632}, [\href{https://arxiv.org/abs/hep-ph/9301212}{{\ttfamily hep-ph/9301212}}].

\bibitem{Bigi:2000wn}
I.~I.~Y. Bigi and N.~G. Uraltsev, \emph{{D0 - anti-D0 oscillations as a probe of quark hadron duality}}, \href{https://doi.org/10.1016/S0550-3213(00)00604-0}{\emph{Nucl. Phys. B} {\bfseries 592} (2001) 92--106}, [\href{https://arxiv.org/abs/hep-ph/0005089}{{\ttfamily hep-ph/0005089}}].

\bibitem{pascual2014qcd}
P.~Pascual and R.~Tarrach, \emph{QCD: Renormalization for the Practitioner}.
\newblock Lecture Notes in Physics. Springer Berlin Heidelberg, 2014.

\bibitem{Mikhailov:1986be}
S.~V. Mikhailov and A.~V. Radyushkin, \emph{{Nonlocal Condensates and {QCD} Sum Rules for Pion Wave Function}}, {\emph{JETP Lett.} {\bfseries 43} (1986) 712}.

\bibitem{Gromes:1982su}
D.~Gromes, \emph{{Space-time Dependence of the Gluon Condensate Correlation Function and Quarkonium Spectra}}, \href{https://doi.org/10.1016/0370-2693(82)90397-5}{\emph{Phys. Lett. B} {\bfseries 115} (1982) 482--486}.

\bibitem{Generalis:1983hb}
S.~C. Generalis and D.~J. Broadhurst, \emph{{The Heavy Quark Expansion and {QCD} Sum Rules for Light Quarks}}, \href{https://doi.org/10.1016/0370-2693(84)90040-6}{\emph{Phys. Lett. B} {\bfseries 139} (1984) 85--89}.

\bibitem{Yndurain:1989hp}
F.~J. Yndurain, \emph{{Nonperturbative Propagators for Scalars and Fermions to All Orders in Their Masses}}, \href{https://doi.org/10.1007/BF01557672}{\emph{Z. Phys. C} {\bfseries 42} (1989) 653--656}.

\bibitem{Bagan:1985zp}
E.~Bagan, J.~I. Latorre and P.~Pascual, \emph{{Heavy and Heavy to Light Quark Expansions}}, \href{https://doi.org/10.1007/BF01441349}{\emph{Z. Phys. C} {\bfseries 32} (1986) 43}.

\bibitem{Bagan:1993by}
E.~Bagan, M.~R. Ahmady, V.~Elias and T.~G. Steele, \emph{{Full contributions of three-dimension and four-dimension QCD vacuum condensates to current current correlation functions and Feynman amplitudes}}, \href{https://doi.org/10.1016/0370-2693(93)91120-C}{\emph{Phys. Lett. B} {\bfseries 305} (1993) 151--156}.

\bibitem{Jamin:1992se}
M.~Jamin and M.~Munz, \emph{{Current correlators to all orders in the quark masses}}, \href{https://doi.org/10.1007/BF01560056}{\emph{Z. Phys. C} {\bfseries 60} (1993) 569--578}, [\href{https://arxiv.org/abs/hep-ph/9208201}{{\ttfamily hep-ph/9208201}}].

\bibitem{Bobrowski:2012jf}
M.~Bobrowski, A.~Lenz and T.~Rauh, \emph{{Short distance D-Dbar mixing}},  in \emph{{5th International Workshop on Charm Physics}}, 8, 2012, \href{https://arxiv.org/abs/1208.6438}{{\ttfamily 1208.6438}}.

\bibitem{Grozin:1994hd}
A.~G. Grozin, \emph{{Methods of calculation of higher power corrections in QCD}}, \href{https://doi.org/10.1142/S0217751X95001674}{\emph{Int. J. Mod. Phys. A} {\bfseries 10} (1995) 3497--3529}, [\href{https://arxiv.org/abs/hep-ph/9412238}{{\ttfamily hep-ph/9412238}}].

\bibitem{mikhailov_nonlocalQCD1992}
S.~V. Mikhailov and A.~V. Radyushkin, \emph{Nonlocal condensates and qcd sum rules for the pion wave function}, \href{https://doi.org/10.1103/PhysRevD.45.1754}{\emph{Phys. Rev. D} {\bfseries 45} (Mar, 1992) 1754--1759}.

\bibitem{Mikhailov:1988nz}
S.~V. Mikhailov and A.~V. Radyushkin, \emph{{Quark Condensate Nonlocality and Pion Wave Function in QCD: General Formalism}}, {\emph{Yad. Fiz.} {\bfseries 49} (1988) 794}.

\bibitem{Braun:1994jq}
V.~Braun, P.~Gornicki and L.~Mankiewicz, \emph{{Ioffe - time distributions instead of parton momentum distributions in description of deep inelastic scattering}}, \href{https://doi.org/10.1103/PhysRevD.51.6036}{\emph{Phys. Rev. D} {\bfseries 51} (1995) 6036--6051}, [\href{https://arxiv.org/abs/hep-ph/9410318}{{\ttfamily hep-ph/9410318}}].

\bibitem{Braun:2003wx}
V.~M. Braun, D.~Y. Ivanov and G.~P. Korchemsky, \emph{{The B meson distribution amplitude in QCD}}, \href{https://doi.org/10.1103/PhysRevD.69.034014}{\emph{Phys. Rev. D} {\bfseries 69} (2004) 034014}, [\href{https://arxiv.org/abs/hep-ph/0309330}{{\ttfamily hep-ph/0309330}}].

\bibitem{Dulibicetal}
L.~Dulibi\'c, B.~Meli\'c and A.~Petrov, \emph{Nonlocal condensate contributions to charm mixing},  \href{https://arxiv.org/abs/in preparation}{{\ttfamily in preparation}}.

\end{thebibliography}\endgroup
\bibliographystyle{JHEP.bst}

\end{document}